\documentclass[pra,aps,groupedaddress,twocolumn,floatfix,nofootinbib]{revtex4}
\usepackage{graphicx}
\usepackage{soul}

\newcommand{\beq}{\begin{equation}}
\newcommand{\eeq}{\end{equation}}
\newcommand{\beqa}{\begin{eqnarray}}
\newcommand{\eeqa}{\end{eqnarray}}

\def\half{\frac{1}{2}}
\def\opone{\leavevmode\hbox{\small1\normalsize\kern-.33em1}}

\begin{document}

\title{Indeterminism in Physics, Classical Chaos and Bohmian Mechanics.\\ Are Real Numbers Really Real?}
\author{Nicolas Gisin \\
\it \small   Group of Applied Physics, University of Geneva, 1211 Geneva 4,    Switzerland}

\date{\small \today}
\begin{abstract}
It is usual to identify initial conditions of classical dynamical systems with mathematical real numbers. However, almost all real numbers contain an infinite amount of information. I argue that a finite volume of space can't contain more than a finite amount of information, hence that the mathematical real numbers are not physically relevant. Moreover, a better terminology for the so-called real numbers is ``random numbers'', as their series of bits are truly random. I propose an alternative classical mechanics, which is empirically equivalent to classical mechanics, but uses only finite-information numbers. This alternative classical mechanics is non-deterministic, despite the use of deterministic equations, in a way similar to quantum theory. Interestingly, both alternative classical mechanics and quantum theories can be supplemented by additional variables in such a way that the supplemented theory is deterministic. Most physicists straightforwardly supplement classical theory with real numbers to which they attribute physical existence, while most physicists reject Bohmian mechanics as supplemented quantum theory, arguing that Bohmian positions have no physical reality. 
\end{abstract}
\maketitle

\section{Introduction}\label{intro}
Physics is often presented as the example of a deterministic explanation of our world. Furthermore, it is often claimed that all good explanations must follow that structure. This is usually illustrated by classical physics, a theory whose explanatory power is truly impressive, despite (or because?) the fact that its limits are well understood. Indeed, the domain of validity of classical mechanics is limited by relativity and quantum theory whose predictions are more accurate when speed and size (or action) get close to critical values determined by the universal constants $c$ and $\hbar$, respectively. 

Classical mechanics is a set of dynamical equations, with initial conditions - typically position and momentum of point particles - given by real numbers. Except for particular cases\footnote{Like, e.g., Norton's dome \cite{NortonDome} and frontal collisions of point particles. \label{FN1}}, these dynamical equations together with the initial conditions determine completely and uniquely the solutions at all future and past times. Hence, the conclusion that classical physics is deterministic.

This has huge consequences. First, as said, this is often taken as the goal of all good scientific explanations. For example, many philosophers and physicists try to formulate quantum physics in such a way as to recover something like classical determinism, despite quantum randomness; in sections \ref{suppvar} and \ref{suppvar2} I discuss Bohmian mechanics in this context. Second, if scientific determinism would be the only good scientific explanation, then it would be highly tempting to conclude that everything covered - at least in principle - by science happens by necessity, i.e. is determined since the big-bang, including all physiological processes.

In my opinion - but this paper is independent of this opinion - this has dreadful consequences: 
our world would be like a movie in a closed box without any spectator. 
If this paper is valid, then there is a greater harmony between physics and our experience \cite{DolevHarmony}.

In the first part of this paper I argue that there is another theory, similar but different from classical mechanics, with precisely the same set of predictions, though this alternative theory is indeterministic\footnote{Here indeterministic is merely the negation of deterministic, i.e. synonymous to non-deterministic: given the present and the laws of nature, there is more than one possible future.}. In a nutshell, this alternative theory keeps the same dynamical equations as classical mechanics, but all parameters, including the initial conditions are given by numbers containing only a finite amount of information. I do not make any metaphysical claims about space, time nor numbers, but notice that the mathematics used in practice is always finite. In sections \ref{realNb}-\ref{randomNb} I argue that this alternative classical mechanics is more natural because it doesn't assume the existence of inaccessible information. One way to argue in favour of limiting physics to numbers with finite information is that any finite volume of space can contain only a finite amount of information (see section \ref{finiteInfo}). Consequently, the huge empirical evidence for classical mechanics equally applies to the alternative indeterministic theory. The alternative theory has the same (enormous) explanatory power, section \ref{ndcm}. It is thus not correct to claim that the empirical evidence and the explanatory power of classical mechanics supports a deterministic world view, since the same body of evidence equally supports an empirically equivalent but indeterministic alternative classical mechanics theory.

In the second part of this paper I argue that every indeterministic theory can be supplemented by additional variables in such a way to render it deterministic (in much the same way as is done by Bohmian mechanics). In brief, it suffices to assume that all the indeterminism that is required at some point in time when, according to the indeterministic theory, God plays dice, i.e. when potentialities becoming actual, could be hidden as supplementary variables in the initial condition of the equivalent deterministic theory, i.e. God played all dice at the big-bang. This closes the circle: deterministic theories are equivalent to indeterministic alternative theories in which real numbers are replaced by finite-information numbers\footnote{As explained in sections \ref{realNb} and \ref{finiteInfo}, see also Fig. 1, finite information numbers contains all computable numbers, but are not restricted to them.}, and indeterministic theories can be supplemented by additional hidden variables in such a way that the supplemented theories are deterministic. 

In sections \ref{suppvar} and \ref{suppvar2} the above rule to supplement indeterministic theories is illustrated on the alternative classical mechanics theory and on standard quantum theory, leading to standard classical mechanics and to Bohmian mechanics, respectively. Admittedly, in these two examples, the supplemented deterministic theories have, in addition to determinism, some elegance which speaks in their favour. However, one may conclude that determinism is too high a price to pay to accept these supplementary hidden variables. Indeed, indeterminism explains nicely, among other things, why probabilistic tools are so powerful in statistical mechanics \cite{Drossel}. Moreover, indeterminism opens the future, makes potentialities a real mode of existence and describes the passage of time when potentialities become actual \cite{NortonTimePasses,Dolev18}.

\section{Classical dynamical systems}\label{cds}
The simplest and thus best known classical dynamical systems are clocks, harmonic oscillators, two bodies interacting via gravity (e.g. one lonely planet orbiting its sun) and similar systems. For such systems, the trajectories are ellipses\footnote{Trajectories whose coordinates are sinuses and cosins, functions that every computer ``knows'' how to calculate efficiently.} (in ordinary or in configuration space), including the cases of degenerate ellipses, i.e. circles and straight lines\footnote{Or the trajectories escape to infinity following parabolas or hyperbolas.}. Such simple dynamical systems are called integrable. They are characterized by their stability: the solution at any time depends only on the leading digits of the initial condition. More precisely, the solution up to any precision $\epsilon$ depends only on the initial condition up to a precision $\epsilon$. Hence, for such simple systems, the far away digits, let's say from the billionth digits on, are physically irrelevant, i.e. don't represent anything physical; rhetorically, I sometimes write that these far away digits have no physical existence or are not physically real. 

However, the fact is that almost all classical dynamical systems are not integrable, they are not simple, but on the contrary are chaotic. In this paper, for clarity, I consider one typical chaotic dynamical system, but it is important to realize that all non-simple classical dynamical systems share the essential features of our example. In this example, we don't consider the solution at all times, but only at a discrete set of times, let's say every microsecond. Furthermore, we assume the system is constrained to remain within the unit interval $[0..1]$, i.e. its coordinate $x$ lies between 0 and 1. Accordingly, its coordinate can be written in binary form as a number like:
\beq\label{x0}
x=0.b_1b_2b_3...b_n...
\eeq
where the $b_j$'s are the bits of $x$ in binary representation (equivalent to the digits in base 10). The dynamics for each time step of this example is given by the following map:
\beq
x\rightarrow\left\{
\begin{array}{ll}
2x & if\ x<\half \\
2x-1 & if\ x\ge\half \\
\end{array}\right.
\eeq
Such a dynamical map is very simple to represent when the coordinate $x$ is represented in binary form:
\beqa\label{map}
x&=&0.b_1b_2b_3...b_n... \nonumber\\
&\rightarrow& 0.b_2b_3...b_n...
\eeqa
At each time-step the bits merely get shifted to the left by one place and the initial leading bit $b_1$ drops out. After $n$ time-steps, the bits shift by $n$ steps to the left. This example of a generic chaotic system is inspired by the baker's map \cite{baker}, though in our example there is a discontinuity at $x=\half$. Note however, that the continuous baker's map shares all features of our map essential for our arguments.

Notice that whether the $x$ in eq. (\ref{x0}) lies in the first half of the unit interval or in the second half is entirely determined by the leading bit $b_1$. This has the following important consequence: whether the system lies within the first half or the second half of the unit interval after $n$ time-steps is determined by the first bit after $n$ time-steps, hence it depends on the $n$th bit of the initial condition.

Such a chaotic system illustrates, for example, the challenge of weather predictions. Let's say that when the system's coordinate $x$ lies on the left of the unit interval, this represents rainy weather, while an $x$ on the right-hand side represents sunny weather. Then, the weather in a week's time depends on infinitesimal bits, e.g. the billionth bit, of the initial condition.

The question here is not whether this billionth bit can be measured, but rather whether this billionth bit has any physical relevance. Clearly, if the initial condition is defined by a real number, then this billionth bit is mathematically well defined. Hence, the question is whether mathematical real numbers are physically real.

\section{Real numbers are are not really real}\label{realNb}
The set of all real numbers is equivalent (isomorphic) to the set of real numbers within the unit interval. Hence, all relevant numbers can be written as in (\ref{x0}), with infinitely many bits $b_j$. This way of writing down real numbers already illustrates the fact that real numbers do, in general, contain an unlimited amount of (Shannon) information, i.e. infinitely many bits. The only exceptions are when the series of bits $b_j$ terminates, or more precisely when all bits after a finite coordinate $m$ are nil: $b_j=0$ for all $j>m$, or when after a finite position $m$ the series of bits repeats itself forever, like, e.g., $0.0111011001010101010101010101...$ which continues with an endless repetition of the pattern $01$, or more generally when there is a finite formula (algorithm) to compute all bits.

Another nice way to illustrate the infinite amount of information in typical real numbers is due to Emile Borel, as nicely told by Gregory Chaitin \cite{Chaitin}. They emphasize that one single real number can contain the answers to all (binary) questions one can formulate in any human language. To see this it suffices to realize that there are only finitely many languages, each with finitely many symbols. Hence,  one can binarize this list of symbols (as routinely done in today's computers) and list all sequences of symbols, first the sequences containing only a single symbol, next those containing two symbols, and so on. This huge list of symbols can then be considered as the bits of a real number. Let's leave 2 bits, $b_n^1b_n^2$, in-between each sequence $S_n$ of symbols:
\beq
0.S_1b_1^1b_1^2S_2b_2^1b_2^2S_3b_3^1b_3^2...S_nb_n^1b_n^2...
\eeq
When the sequence $S_n$ of symbols doesn't represent a binary question, we set these two bits to 0 ($b_n^1b_n^2=00$). When they represent a question whose answer is {\it yes}, we set these bits to 01 and if the answer is {\it no} we set them to 10. This procedure is not efficient at all, but who cares\footnote{An already more economical coding would be to ignore all sequences of symbols that do not represent any question and add after each meaningful question a single bit coding for the answer.}: since a real number has infinitely many bits, there is no need to save space! Hence, one can really code the answers to all possible (binary) questions in one single real number. This illustrates the absurdly unlimited amount of information that real numbers contain. Real numbers are monsters!

In the next section I argue that a finite volume of space can contain no more than a finite amount of information. Following this reasonable assumption, I argue that the so-called real numbers are not really real. More precisely, I argue that the mathematical real numbers are not physically real, by which I mean that they do not represent anything physical. Indeed, the thesis of this paper is that all of physics can be done with only finite-information numbers. Note that these numbers contain all computable numbers, but are not restricted to them: they also contain "numbers" whose far away digits are undetermined, i.e. not yet determined\footnote{This is somewhat reminiscent of Brouwer's indeterminate numbers \cite{IndeterminateNumbersPosy}.}, as illustrated in Fig. 1. All numbers containing an infinite amount of information, cannot represent physical entities; specifically, they cannot be used, and in fact are not used, to describe initial conditions\footnote{To be clear  - I am not making claims pertaining to the nature of numbers, or making assumptions regarding the reality or unreality of numbers. My concern is with distinguishing numbers that have physical significance from those that do not. Nor am I at all concerned with the well known and daunting task of accounting for the applicability of math in science. Finally, I am not arguing against the use of real numbers as a useful tool for calculus, simply only finite-information number can represent something physical.}, see also \cite{DowekRealNb13}. Moreover, in practice, one never uses real numbers, except to prove some general abstract non-constructive existence theorems. The fact that one doesn't need real numbers in practice is quite obvious, as one never accesses an infinite amount of information. Furthermore, today all predictions can be - and most of the time are - encoded in computers, computers that obviously hold at most a finite amount of bits, as emphasized in the next section. Consequently, physics is actually done using only finite-information numbers and, as we'll see in section \ref{ndcm}, classical physics with finite-information initial condition is a well defined indeterministic alternative theory to classical mechanics. Admittedly, one may prefer to postulate that real numbers are physically significant, as I discuss in section \ref{suppvar2}.

\section{A finite volume in space contains at most finite information}\label{finiteInfo}
Here I present an argument supporting the claim that real numbers cannot represent anything physical. This argument is based on the assumption that no finite volume of space can contain an infinite amount of information. This is a well accepted result that follows from the holographic principle, known as the Bekenstein bound \cite{HoloPrinciple,BekensteinBound}. In brief, any storage of a bit of information requires some energy and large enough energy densities trigger black holes. However, for the purpose of my argument, I believe a much simpler reasoning suffice to convince oneself that every bit of information occupies some space, hence that information density is limited. Let me now present this reasoning based on an intuitive assumption.

The enormous progress in information storage over the last few decades profoundly impacts our society\footnote{Allow me a side remark. The enormous progress in information storage and the relatively poor progress in energy storage explains why the science fiction of half a century ago completely missed the ``internet revolution''. In the science fiction of those days no one could hold an encyclopedia in his pocket, but everyone was flying thanks to small backpacks.}. Today, everyone knows that they hold gigabytes of information in their pockets and that companies like Google and agencies like the NSA store everything that transits through the internet. Furthermore, everyone knows also that each stored bit requires some space. Not much, possibly soon only a few cubic nanometre ($10^{-18}$ $mm^3$), but definitively some finite volume. Consequently, assuming that information has always to be encoded in some physical stuff, a finite volume of space cannot contain more than a finite amount of information. At least, this is a very reasonable assumption.

Here I would like to explore the deep conceptual consequences of this assumption, an assumption easy to formulate and defend in our information-based society, but an assumption hardly conceivable a century ago. Remember indeed, that the modern concept of information was formalized by Shannon only in the 1940's.

Consider a small volume, a cubic centimetre let's say, containing a marble ball. This small volume can contain but a finite amount of information. Hence, the centre of mass of this marble ball can't be a real number (and even less 3 real numbers), since real numbers contain - with probability one - an infinite amount of information. Classical physics describes the centre of mass of the ball by 3 real numbers; and this is an extremely efficient description. But the assumption that a finite volume of space can't contain more but a finite amount of information implies that the centre of mass of any object cannot be identified with mathematical real numbers. Real numbers are useful tools, but are only tools. They do not represent physical reality. 

Admittedly, according to today's physics the above argument is a bit misleading, since we know that, ultimately, the marble ball and its centre of mass should be described by quantum physics, including quantum indeterminacy (often called uncertainty). This is correct, of course. But let's continue with classical physics because, first, it remains extremely useful today, and, secondly, it is often presented as the archetype of deterministic theories. The main point of this paper is that classical physics is deterministic only if one attributes to the tool of real numbers physical significance. As soon as one realizes that the mathematical real numbers are ``not really real'', i.e. have no physical significance, then one concludes that classical physics is not deterministic, as we elaborate in section \ref{ndcm}. Actually, things are even worse, as we explain in the next section.

\section{Mathematical real numbers are physical random numbers}\label{randomNb}
Some real numbers can be computed up to arbitrary precision with a computer, like for example all rational numbers and numbers like $\sqrt{2}$ and $\pi$. Such computable numbers contain only finite information, the length in bits of the shortest program that outputs their bits. Note that since there are only countably many programs, the set of real numbers that can be calculated by a computer is infinitely smaller than the set of all real numbers. More precisely, the set of computable numbers is countably-infinite, like the integers, while the set of real numbers is continuously-infinite, like the mathematical points on a line. Consequently, real numbers are uncomputable with probability one. Or, equivalently, the set of computable numbers has measure zero among the set of all real numbers. For more see \cite{Chaitin}.

\begin{figure}[h]
\includegraphics[width=6cm]{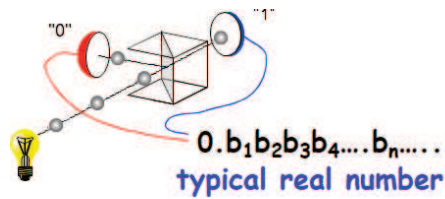}
\caption{\it The series of bits of typical of real numbers is, after some initial finite bit string, indistinguishable from a truly random sequence of bits, as produce by a physical random number generator, here illustrated by a quantum random number generator (single photons on a beam-splitter followed by two single-photon detectors). Hence, there are two ways one may think of typical real numbers. First, one may think of them as dynamical quantities: the far away bits don't have yet any determined value, the values are continuously produced as time passes. Second, one may think of real numbers with all their bits given at ones, as if their random values were produced at some initial time. The second view is the standard view, while the first one is the view advocated here. Note that computable numbers correspond to pseudo-random number generators where all the information lies in the finite seed. The finite information numbers used here include both the computational numbers and the dynamical real numbers whose far away bits are not yet determined.}
\end{figure} 

The above simple observation has the following important consequence: after the first bits, the next bits of almost all real numbers are random: they don't follow any structure. These bits are as random as the outcome of quantum measurements (on half a singlet, let's say), i.e. they are as random as possible\footnote{Some caution is due here, as not all bits of all real numbers are random, as illustrated by the following example. Define a number with all bits at even positions identical to the corresponding bits of the computable number $\pi$ and all bits at odd positions determined by the successive outcomes of a given (infinite) sequence of ``true coin tosses'', e.g. the outcomes of a quantum random number generator. In such an artificial case, every second bit is predictable, but all others are truly random.\label{randomPi}} \cite{Chaitin}. Actually, one can't even name or characterize real numbers, as there are only countably many names and characterizations. Hence, almost all real numbers are totally outside our grasp: we can't say anything about them, except that their digits are random, they have no structure. Indeed, if the digits of a real number had some structure, this very structure would allow one to characterise and name that number.

Accordingly, to name them ``real number'' is seriously confusing. A better terminology would be to call them ``random numbers''. Unfortunately, Descartes named them ``real'' to contrast them with the complex numbers, those numbers that include the square root of $-1$, traditionally denote $i$. Hence:

Mathematical real numbers are physical random numbers.

I think it can be speculated that had we learned in school to name such numbers "random" rather than "real", we would be less inclined to adopt a deterministic outlook on the basis of the science they figure in.

\section{Non-deterministic classical physics}\label{ndcm}
In this section we return to physics, now that we have established that the ``mathematical real numbers are physical random numbers'', i.e. most mathematical real numbers are physically immaterial. We define alternative classical mechanics as the same set of dynamical equations as standard classical mechanics, but all parameters, notably the initial conditions, are given by finite-information numbers. Note that for integrable dynamical systems, the coordinates at all times given by a finite-information parameter $t$ are themselves described by finite-information numbers. Hence, all predictions about integrable systems remain unchanged.

However, looking back to the example of the typical classical chaotic dynamical system of section \ref{cds}, we recall that the leading bit describing the system, i.e. the bit that determines whether the system lies of the left half (rainy weather) or on the right half of the unit interval (sunny weather) after $n$ time-steps, depends on the $n$th bit of the initial condition, $b_n$. But, if $n$ is large enough, this $n$th bit of the initial condition had, at the time corresponding to the initial condition, no physical significance. Hence, according to our alternative classical mechanics, chaotic dynamical systems are truly random. Let me emphasize that they are not merely random for all practical purposes, but that they are truly random, as random as quantum measurement results. This randomness has nothing to do with technological limitations, it is intrinsic pure randomness.

The view I am suggesting is that the first bits in the expression of $x$ are ``really real'' (e.g., at present, it is really rather sunny or rather rainy), while the very far away bits are totally random. As time passes they get shifted to the left, one position at each time-step\footnote{The time-steps are used here only as illustration. Time could pass much smoother, with the propensities of all bits varying slowly.}. Hence, step by step they acquire some definite value. As time passes they have a changing disposition (or propensity) \cite{Dorato11} to hold their eventual value. This propensity changes at each time-step, similarly to the quantum probability of physical quantities of quantum systems that also evolve as time passes. I suggest that this is similar to the Brownian motion of some particle that evolves in-between two sticky plates (that code for the bit values 0 and 1) until it eventually sticks either on the left plate or on the right plate\footnote{Note that this is similar to the quantum state of a quantum-bit (qubit) in spontaneous localization models described with stochastic Schr\"odinger equations \cite{QSD}.}. Accordingly, the openness of the future enters gradually, sometimes on millisecond scales and for other systems on scales of millions of years. This is just a brief statement of an idea, an elaboration will follow in a future paper.

One may object that this view is arbitrary as there is no natural bit number where the transition from determined to random bits takes place. This is correct, though not important in practice as long as this transition is far away down the bit series. The lack of a natural transition is due to the fact that, in classical physics, there is no equivalent to the Plank constant of quantum theory. But this is quite natural, as the fact is that when one looks for this transition in the physical description of classical systems, one hits quantum physics. 

Admittedly, Newton's equations, as well as Maxwell's equations, are deterministic: given initial conditions in the form of real numbers, all the future and past are fixed\footnote{Up to exceptional cases, see footnote \ref{FN1}.}. But the fact that these equations are mathematically deterministic doesn't imply that physics is deterministic. For example, this is definitively not the case when the initial conditions of chaotic systems are not identifiable with mathematical real numbers, as in our alternative classical mechanics. Consequently, whether classical physics is deterministic or not is not a scientific question, but depends on the physical significance one associates with mathematical real numbers.

As the philosopher Elizabeth Anscombe emphasized \cite{Anscombe}, {\it ``the high success of Newton’s astronomy was in one way an intellectual disaster: it produced an illusion (...) for this gave the impression that we had here an ideal of scientific explanation; whereas the truth was, it was mere obligingness on the part of the solar system, by having had so peaceful a history in recorded time, to provide such a model''}.

\section{Supplementary variables...}\label{suppvar}
So far we have seen that physics is non-deterministic and that this is true both of quantum \cite{GisinQchance14} and alternative classical mechanics. In this section we turn to the natural question of whether one could add supplementary variables to quantum and to alternative classical mechanics in order to restore determinism. 

That it is possible to do so in the case of classical mechanics is well known. It suffices to add the mathematical real numbers, as is usually done without even mentioning that these are supplementary variables. Once these real numbers are added and postulated to be part of the ontology of the theory, the so extended theory is deterministic. Somehow, all the randomness has been pushed back to the (unattainable) initial conditions, as discussed in the next section. 

For physicists this may look like a joke: we first argued at length that "`real numbers are not really real"' just to next introduce them again. But notice what is achieved by viewing things in this way. The real numbers are certainly not necessarily part of the ontology of classical physics, it is not the experimental facts that force physics to include real numbers in the ontology of classical physics. Hence, at first, classical physics is non-deterministic. However, all non-deterministic theories can be turned deterministic by adding supplementary variables. In full generality, it suffices to add, for example, as supplementary variables all results of all future measurements, while making sure that these supplementary variables remain hidden as long as the corresponding measurements haven't happen. In fact, that's exactly how classical mechanics is done: postulating that the initial condition of all classical dynamical systems are faithfully described by real numbers is an elegant way of adding all future results, while making sure that they remain inaccessible for long enough a time. Admittedly, just adding future results and postulating that there are inaccessible won't convince any scientist. Adding the real numbers to classical physics is much more convincing, because it is elegant. But is it truly different? 

And what about quantum physics? Here there is a well known way to add supplementary variables in such a way as to turn quantum mechanics deterministic. This is known as Bohmian mechanics (or the de Broglie-Bohm pilot wave) \cite{Bohm52,BellBohm,DurrTeufel}. Essentially, one postulates that all particles always have well defined - though inaccessible - positions and that, at the end of the day, all measurements are positions measurements (position of some pointer, position of some electrons that turn on/off some LEDs, etc). These particle positions, which I name Bohmian positions, are guided by the solution of the usual Schr\"odinger equation in a clever way such that if the initial positions are assumed to be statistically distributed according to the usual quantum probabilities, i.e. $|\psi(\vec x)|^2$, then the statistical distribution of the Bohmian positions remain in accordance with quantum probabilities at all times. This is very elegant and, like for the real numbers for classical physics, adding Bohmian positions to the ontology of the theory turns quantum physics into a deterministic theory. Note that it requires also to ``trust'' real numbers, as the Bohmian positions and the quantum state vector use them. 

A point of caution is due here: In case of systems composed of more than one particle, one should realize that the evolution of any particle, let's say the first one, depends on the entire wave-function. Hence, it depends also on what happens at the location of the other particles, i.e. each particle is guided in a non-local way. This is necessary because quantum predictions violate the Bell inequality, hence all alternative (or supplemented) theories that reproduce quantum predictions must contain some non-local features \cite{Bellspeakable, BrunnerRMP14, GisinQchance14}. But this is probably why most physicists reject Bohmian mechanics: they dislike explicit (though unavoidable) non-locality.

\section{... push randomness back to the initial condition}\label{suppvar2}
As we saw in the previous section, both alternative classical and quantum non-determinism can be turned deterministic by adding supplementary variables. In both cases, the complemented deterministic theory is rather elegant. In both cases, the original randomness is pushed back to the initial conditions. 
Indeed, as time passes, instead of new bits in the series (\ref{x0}) gaining determined values, new bits from the initial condition gain relevance. Hence, we face a choice: either the fact that at present certain things happen and others do not is interpreted as revealing, retroactively, information about long past initial conditions, or else, we understand the present as the result of indeterminate reality, and the future as open. If we care about how we experience reality, the later option is obviously superior.   

Noteworthy, it is a fact that almost all physicists do complement alternative classical mechanics with the mathematical real numbers; they do so even without thinking about it. At the same time, almost all physicist reject Bohmian mechanics arguing that it is unnecessarily complicated and doesn't lead to new physics. However, one may argue that the real numbers accepted in classical physics are also unnecessarily complicated (remember, they contain an infinite amount of information. Isn't that hugely complex?). Furthermore, one can ask which new physics the real numbers produced.

\section{conclusion}
In our society the concept of information is ubiquitous. Today, it is quite natural to assume that no finite volume of space can hold more but a finite amount of (Shannon) information, as measured by bits. Consequently, I argue that one should not attribute to real numbers, i.e. to numbers that contain an infinite amount of information, any physical significance. This observation implies that there is a simple alternative to standard classical mechanics, based on finite-information numbers, which is a non-deterministic theory although it has exactly the same predictive and explanatory powers.

At the time of Laplace the concept of information, in particular its quantization in terms of bits, was non-existing. Hence, it was natural to identify initial conditions of classical dynamical systems with the mathematical real numbers. But today that we know that ``real numbers'' contain an infinite amount of information and, as suggested above, that they would be better called ``random numbers'', we should realize that such numbers can't be the basis for determinism.

Accordingly, both classical and quantum theories can and, I claim, ought to be regarded as non-deterministic. Of course, one may want to complement these theories with supplementary variables in such a way that the complemented theory is deterministic. Note that this can be done both for quantum and for classical physics, as seen in sections \ref{suppvar} and \ref{suppvar2}; in both cases the supplementary variables are inaccessible. The fact is that most physicists easily complement classical physics, but are reluctant to make the similar move for quantum physics. 

In summary, physics with all its predictive and explanatory powers can well be presented as intrinsically non-deterministic. The dominant view according to which classical physics is deterministic is due, first, to a false impression generated by it's huge success in astronomy and in the design of clocks and other simple mechanical (integrable) systems, and, second, to a lack of appreciation of its implication for (infinite) information density. 

Finally, an indeterministic world is hospitable to Res Potentia and to the passage of time \cite{GisinTimePasses,past,DolevPassage}. \\

\small
\section*{Acknowledgment} This work profited from stimulating discussions with Augustin Baas, Cyril Branciard, Barbara Drossel, Florian Fr\"owis, Michael Hall, John Norton, Valerio Scarani and Christian W\"uthrich. Financial support by the European ERC-AG MEC is gratefully acknowledged.\\ \\


\begin{thebibliography}{99}
\bibitem{past} Admittedly, the non-determinism of our alternative classical mechanics questions when do potentialities become actual. Continuously as part of the dynamics \cite{CollapseWhatElse}? This might shed new light on the old and infamous quantum ``measurement'' problem. Also, questions about time reversal may get interesting answers. If one inverts the map (\ref{map}), new leading bits must be added at each time-steps, but without losing any bit. Hence, both the future and the past are open, but the present contains more information about the past than about the future.
\end{thebibliography}
\end{document}